\documentclass[cameraready]{Interspeech}

\title{Robust LLM-based Audio-Visual Speech Recognition with Sparse Modality Alignment and Visual Unit-Guided Refinement}

\author[affiliation={1,2}, orcid=0009-0005-6523-5542]{Fei}{Su}
\author[affiliation={1,2}]{Cancan}{Li}
\author[affiliation={2,1}, orcid=0000-0001-9344-7415,correspondingauthor]{Juan}{Liu}
\author[affiliation={4}]{Wei}{Ju}\author[affiliation=4]{Hongbin}{Suo}
\author[affiliation={3,5}, orcid=0000-0002-6406-1983,correspondingauthor]{Ming}{Li}

\address{
    $^1$ School of Computer Science, Wuhan University, China \\
    $^2$ School of Artificial Intelligence, Wuhan University, China \\
    $^3$ School of Artificial Intelligence, The Chinese University of Hong Kong, Shenzhen, China \\ 
    $^4$ AI Center, OPPO, China \\
    $^5$ Digital Innovation Research Center, Duke Kunshan University, China 
}

\email{fei.su@whu.edu.cn, cancan.li@whu.edu.cn, liujuan@whu.edu.cn, juwei@oppo.com, suohongbin@oppo.com, ming.li.cuhksz@gmail.com}

\keywords{audio-visual speech recognition, visual discrete units, large language models}

\usepackage{comment}

\usepackage{amsmath,graphicx,hyperref}
\usepackage{amssymb,mathtools,bm}
\usepackage{microtype}
\usepackage{enumitem}

\usepackage{booktabs}   
\usepackage{makecell}   
\usepackage[table]{xcolor} 

\usepackage{multirow}
\usepackage{array}

\usepackage{pifont}
\newcommand{\cmark}{\ding{51}}
\newcommand{\xmark}{\ding{55}}

\usepackage{url}

\begin{document}
\maketitle

\begin{abstract}
   Audio-Visual Speech Recognition (AVSR) integrates acoustic and visual information to enhance robustness in adverse acoustic conditions. Recent advances in Large Language Models (LLMs) have yielded competitive automatic speech recognition performance and shown effectiveness for AVSR. However, prior approaches project audio and visual features independently or apply shallow fusion, limiting cross-modal alignment and complementary exchange while increasing the LLM's computational load. To address this, we propose AVUR-LLM, an LLM-based Audio-Visual Speech Recognition via Sparse Modality \textbf{A}lignment and \textbf{V}isual \textbf{U}nit-Guided \textbf{R}efinement. Experiments on LRS3 demonstrate state-of-the-art results for AVSR. Under additive-noise conditions at \(0\,\mathrm{dB}\) SNR, it achieves \(37\%\) relative improvement over the baseline system. 
\end{abstract}

\section{Introduction}

Audio-Visual Speech Recognition (AVSR) integrates acoustic signals with visual cues to enhance robustness and accuracy under adverse acoustic conditions. Early supervised pipelines, spanning DNN/CTC~\cite{noda2015audio,graves2006ctc}, sequence-to-sequence~\cite{afouras2018deep}, and RNN-T formulations~\cite{makino2019recurrent}, laid the groundwork but remain vulnerable to noise. Self-supervised learning improves data efficiency and transferability by pretraining strong speech and audio-visual encoders. 
Self-supervised acoustic models such as wav2vec~\cite{baevski2020wav2vec} and WavLM~\cite{chen2022wavlm} enhance representation quality. 
Audio-visual pretraining with AV-HuBERT~\cite{shi2022avhubert} and AV-data2vec~\cite{lian2023av} further boosts the performance of AVSR, and automatic-label training provides additional gains~\cite{ma2023auto}. In recent years, Large language models (LLMs) have been shown effective for speech recognition, with studies integrating them as parametric language models or as multi-modal reasoners alongside speech encoders, achieving competitive recognition performance~\cite{chen2023hyporadise}. In AVSR, recent LLM-based frameworks fuse Whisper~\cite{radford2023robust} with visual features~\cite{rouditchenko2024whisper} and couple Whisper with LLaMA variants to exploit multimodal context~\cite{yeo2024visual,cappellazzo2025scaling}.

Prior studies on LLM-based AVSR~\cite{Cappellazzo2025Large,yeo2025mms,cappellazzo2025adaptive} typically use self-supervised encoders as feature extractors. Acoustic and visual features are temporally compressed and linearly projected to the LLM embedding space, and the LLM is then applied for prediction or shallow fusion. Another approach in LLM-based speech recognition adopts a rescoring strategy~\cite{radhakrishnan2023whispering,chen2024s,liu2025listening,ghosh2024lipger}, providing acoustic features to the LLM or applying text-only rescoring to reduce recognition error. However, despite these advances, LLM-based speech recognition still faces limitations: Tight coupling to acoustic/visual features increases memory consumption and sensitivity to input noise. Furthermore, multimodal fusion lacks fine-grained control. To reduce the computational cost and improve the robustness, recent works~\cite{shi2022learning, kim2024efficient,mousavi2024should} discretize audio or visual embeddings via vector quantization, yielding token sequences that suppress high-frequency variability and shorten LLM prompts. Building on these works, we introduce a lightweight and stable alignment mechanism that enables controlled cross-modal exchange without disturbing the pretrained audio pathway.

Motivated by the above, our work further develops controlled multimodal interaction and adaptive fusion, while additionally employing speech-informative visual discrete tokens to guide LLM-based rescoring. In this paper, we propose AVUR-LLM. The main contributions are: (1) a stability-oriented sparse alignment that inserts lightweight cross-modal blocks into the upper audio encoder layers, (2) a confidence-aware fusion that modulates token-level visual injection, with visual discrete units-based prompt for LLM rescoring, (3) experiments on LRS3 dataset show that our model outperforms AVSR baseline systems and reduces WER under both clean and noisy conditions, achieving a 37\% relative reduction at \(0\,\mathrm{dB}\) SNR.

\begin{figure*}[t]
  \centering
  \vspace{-2pt}
  \includegraphics[width=\textwidth,trim=6 3 6 6,clip]{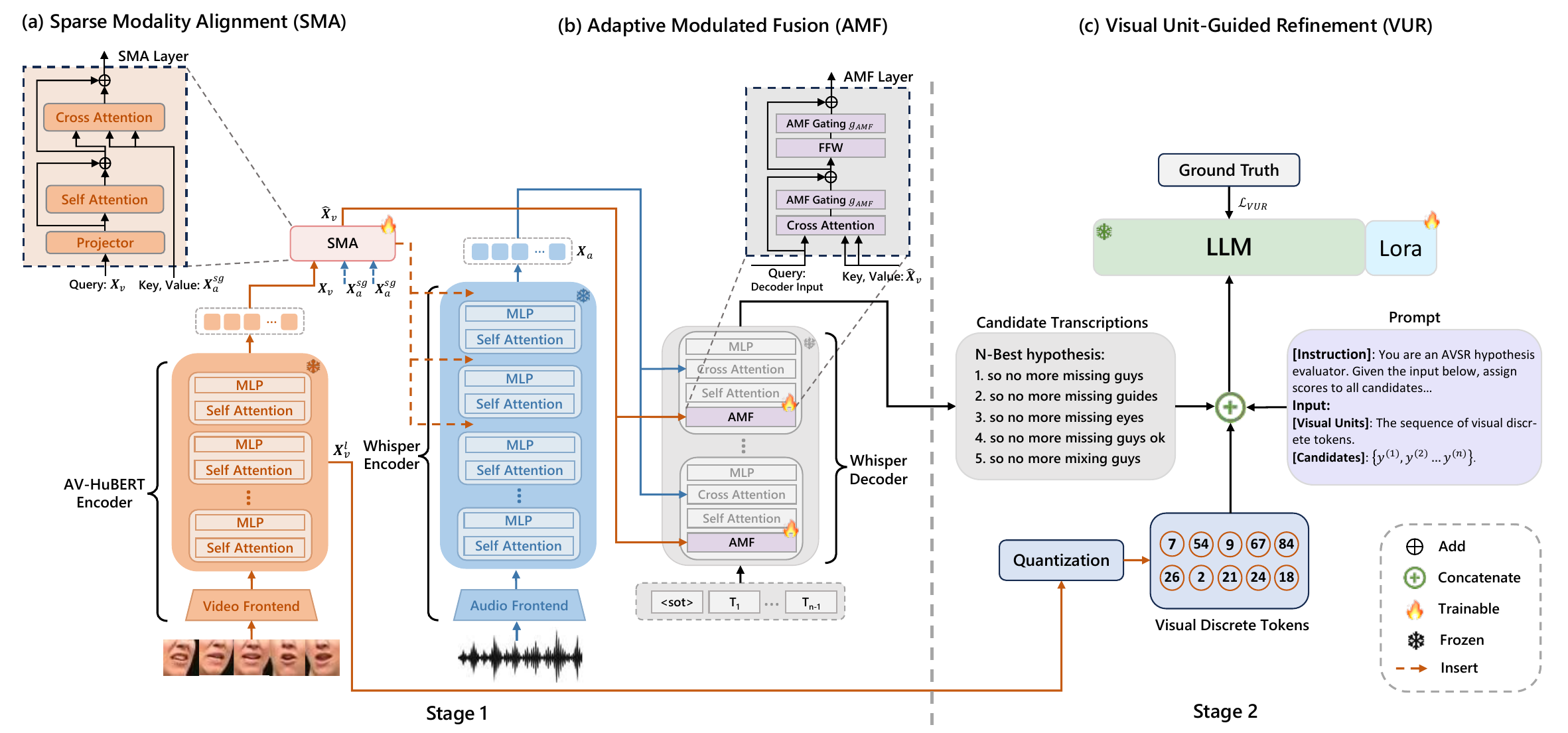}
  \vspace{-4pt}
  \caption{Illustration of AVUR-LLM. 
 (a) The Sparse Modality Alignment (SMA) module, audio-conditioned cross-attention sparsely inserted into the upper audio encoder layers; (b) The Adaptive Modulated Fusion (AMF) module, confidence-aware gating that adaptively modulates visual injection in the decoder; (c) The Visual Unit-Guided Refinement (VUR) module, discretizes mid-layer visual features into tokens for LLM rescoring. Training has two stages: \emph{Stage 1} performs SMA+AMF and outputs $N$-best hypotheses; \emph{Stage 2} extracts and compresses \(X_v^{\ell}\) into visual discrete tokens and uses a LoRA-adapted LLM to rescore the hypotheses. Here \(\mathbf{X}_{a}\) and \(\mathbf{X}_{v}\) denote audio and visual encoder features. \(\hat{\mathbf{X}}_{v}\) denotes the SMA-refined visual features, \(\mathbf{X}_{v}^{\ell}\) denotes visual features extracted from encoder layer \(\ell\). \(\mathbf{X}_{a}^{\mathrm{sg}}\) denotes the audio features serving as keys/values with stop-gradient. \(g_{\mathrm{AMF}}\) denotes the AMF gate coefficients.}
  \label{fig:model}
  \vspace{-8pt}
\end{figure*}

\section{Methodology}

Our proposed framework is illustrated in Fig.~\ref{fig:model}. The methodology involves sparse modality alignment, adaptive modulated fusion and visual unit-guided refinement.

\subsection{Sparse Modality Alignment}
\label{sec:sma}
We propose Sparse Modality Alignment (SMA), which sparsely inserts lightweight alignment blocks into the upper audio encoder. In each block, visual features act as queries and attend to audio context to calibrate the visual representations.

We employ the Whisper encoder~\cite{radford2023robust} for audio and the AV-HuBERT encoder~\cite{shi2022avhubert} for visual input, extracting features $\mathbf{X}_{a} \in \mathbb{R}^{T_a \times D_a}$ and $\mathbf{X}_{v} \in \mathbb{R}^{T_v \times D_v}$, respectively. We then leverage the audio pathway’s higher temporal fidelity and strong pretraining to guide visual alignment and enhancement. Aligning the visual features to the audio temporal resolution mitigates frame-rate mismatch. Restricting alignment to the upper layers of the audio encoder and applying stop-gradient to the audio keys/values preserve acoustic representations, stabilize optimization, and incur modest computational overhead.

Specifically, the audio resolution is preserved, while visual features are upsampled via a lightweight differentiable resampler and projected to the audio feature dimension, yielding \(\mathbf{X}'_{v}=\mathcal{R}_{v}(\mathbf{X}_{v})\,W_{v}+b_{v}\) with \(\mathbf{X}'_{v}\in\mathbb{R}^{T_a\times D_a}\). Here, \(\mathcal{R}_{v}(\cdot)\) denotes lightweight differentiable resampling that aligns the visual features to the temporal resolution \(T_a\). \(W_{v}\in\mathbb{R}^{D_v\times D_a}\) and \(b_{v}\in\mathbb{R}^{D_a}\) are learnable parameters of a linear projector converting the feature dimension from \(D_v\) to \(D_a\).

Subsequently, we insert lightweight blocks into the upper audio encoder layer to align visual features with audio cues while keeping the audio features fixed. Let \(\mathbf{X}_{a}\in\mathbb{R}^{T\times D}\) and \(\mathbf{X}_{v}'\in\mathbb{R}^{T\times D}\), where \(T=T_{a}\) and \(D=D_{a}\). Each block:
\begin{equation}
\label{eq:sab}
\begin{aligned}
\mathbf{X}_{v}^{\mathrm{sa}} &= \mathrm{MHA}(\mathbf{X}_{v}',\,\mathbf{X}_{v}',\,\mathbf{X}_{v}'),\\
\hat{\mathbf{X}}_{v} &= \mathbf{X}_{v}' + \mathrm{MHA}(\mathbf{X}_{v}^{\mathrm{sa}},\,\mathbf{X}_{a}^{\mathrm{sg}},\,\mathbf{X}_{a}^{\mathrm{sg}}),
\end{aligned}
\end{equation} 
 where  \(\mathrm{MHA}\) denotes Multi-Head Attention~\cite{vaswani2017attention}, \(\mathrm{sa}\) denotes self-attended visual features, \(\mathrm{sg}\) denotes stop-gradient, ensuring that audio features act only as keys/values and remain unchanged. We insert three alignment blocks into the upper layers of the audio encoder, so that alignment focuses on high-level representations and preserves computational efficiency and audio encoder stability.

\subsection{Adaptive Modulated Fusion}
\label{sec:amf}

To adaptively exploit complementary visual cues during decoding, we first estimate a token-level acoustic reliability signal via a forward-only probe on the decoder input and audio keys/values. Then, we use this signal to gate the magnitude of visual injection. Inspired by \cite{rouditchenko2024whisper}, we additionally apply a tanh-based direction gate that determines whether the injected visual features operate in a positive or negative direction.

Specifically, we adapt Whisper’s decoder by inserting a forward-only acoustic probe at each layer, with gradients stopped through the queries and audio features. Given decoder queries \(\mathbf{Q}_{\ell}\in\mathbb{R}^{T_y\times D}\) and audio memory \(\mathbf{X}_{a}\in\mathbb{R}^{T\times D}\), the probe attention is:
\begin{equation}
\label{eq:amf_probe}
A_{\ell}
=\mathrm{softmax}\left(
\frac{(\mathbf{Q}_{\ell}^{\mathrm{sg}}W_{Q})\,(\mathbf{X}_{a}^{\mathrm{sg}}W_{K})^{\top}}{\sqrt{D}}
\right).
\end{equation}
The token-wise acoustic uncertainty \(S_{\ell,t}\) is computed as the entropy of the probe attention over audio frames \(A_{\ell}^{t}\), normalized to \([0,1]\) by \(\log T\). The values of \(S_{\ell,t}\) indicate the concentration of attention on audio frames, where smaller values correspond to more reliable acoustic contexts and larger values reflect higher acoustic uncertainty. Therefore, the amplitude of visual feature injection is defined as:
\begin{equation}
\label{eq:amf_gate_linear}
g_{\mathrm{amp},t} \;=\; \sigma\bigl(a\,S_{\ell,t}+b\bigr),
\end{equation}
where $a,b\in\mathbb{R}$ are learnable scalars controlling slope and offset, and $\sigma(\cdot)$ denotes the sigmoid function, ensuring $g_{\mathrm{amp},t}\!\in\!(0,1)$.

Given the decoder input \(\mathbf{x}_t\), the visual context \(\mathbf{c}^{(v)}_t\) is obtained by MHA with query \(\mathrm{LN}(\mathbf{x}_t)\) and keys/values \(\hat{\mathbf{X}}_{v}\). The adaptive modulated fusion (AMF) gating uses two scalars per layer: \(g_{\mathrm{AMF},\mathrm{att}}\) for the cross-attention branch and \(g_{\mathrm{AMF},\mathrm{ff}}\) for the feed-forward branch. Each gate combines a direction term, obtained by applying \(\tanh(\cdot)\) to a learnable scalar, with a token-wise amplitude term \(\sigma(\cdot)\) derived from the acoustic uncertainty \(S_{\ell,t}\). The AMF gates modulate the cross-attention and feed-forward residuals, respectively: 
\begin{equation}
\label{eq:amf_update}
\begin{aligned}
\mathbf{x}'_t &= \mathbf{x}_t + g_{\mathrm{AMF},\mathrm{att}}\,\mathbf{c}^{(v)}_t,\\
\mathbf{y}_t  &= \mathbf{x}'_t + g_{\mathrm{AMF},\mathrm{ff}}\,\mathrm{FFW}\big(\mathrm{LN}(\mathbf{x}'_t)\big),
\end{aligned}
\end{equation}
where LN is Layernorm, and FFW is a Multi-Layer Perceptron (MLP). We train using token-level cross-entropy between the
model’s predicted transcript and the ground-truth tokens. During optimization, the trainable components are the sparse modality alignment module and the adaptive modulated fusion module. At inference, beam search yields an \(N\)-best list \(Y=\{y^{(i)}\}_{i=1}^{N}\) with sequence scores
\(
s_{\mathrm{infer}}\big(y^{(i)}\big).
\)

\subsection{Visual Unit-Guided Refinement}
\label{sec:vur}

Since the AV-HuBERT encoder’s visual representations carry speech-relevant structure, we discretize them into visual tokens and use the resulting sequence as conditioning input to an LLM for $N$-best rescoring. Unlike prior LLM-based approaches~\cite{Cappellazzo2025Large,yeo2025mms,liu2025listening} that feed the LLM with continuous audio/visual feature projections or fused representations, VUR provides compact visual token sequences derived from AV-HuBERT as an explicit visual prompt.

Concretely, we use visual embeddings \(\mathbf{X}_{v}^{\ell}\in\mathbb{R}^{T_v\times D_v}\) from the visual encoder, with frame vectors \(\mathbf{x}_{v,t}^{\ell}\in\mathbb{R}^{D_v}\). An offline \(K\)-means codebook with centroids \(\{\mathbf{k}_1,\dots,\mathbf{k}_K\}\) is learned~\cite{shi2022avhubert}, and each frame is quantized by assigning it to the nearest centroid \(\mathbf{k}_j\). To exploit temporal stationarity and shorten the token stream, we apply run-length compression. Maximal contiguous spans with identical labels are collapsed into a single token by averaging the frames in each span and retaining the common label~\cite{yeo2024visual}.
 Let \(R_1,\dots,R_M\) denote the maximal contiguous index sets with identical labels. For each \(R_m\) we compute the averaged feature \(\bar{\mathbf{x}}_{v,m}^{\ell}=|R_m|^{-1}\sum_{t\in R_m}\mathbf{x}_{v,t}^{\ell}\) and assign the common label \(k_m\). The resulting sequence of visual discrete units is cached once per utterance and reused for \(N\)-best rescoring.

Given the first-stage inference results, a LLM~\cite{touvron2023llama} is adapted with LoRA~\cite{hu2022lora} to assign a scalar score $r_i$ to each candidate, where LoRA parameters are trainable. Let $i_{\mathrm{gt}}$ denote the candidate that is closest to the ground-truth transcript. Training minimizes a list-wise softmax:
\begin{equation}
\label{eq:vugr_softmax}
\mathcal{L}_{\mathrm{VUR}}
= -\log \frac{\exp(r_{i_{\mathrm{gt}}})}{\sum_{j=1}^{N}\exp(r_j)},
\end{equation}
where $r_j\in\mathbb{R}$ is the LLM-assigned score for the $j$-th candidate, $N$ is the size of the $N$-best list.

\section{Experimental Results}
\label{sec:experiments}

\subsection{Experimental Settings}
All experiments are conducted on the LRS3 dataset~\cite{afouras2018lrs3}, which is a widely used publicly available English audio-visual speech recognition dataset. The dataset comprises approximately 433 hours of transcribed TED talk videos. In addition to the full training set, we also evaluate performance on the 30-hour ``trainval'' subset. Additionally, following~\cite{rouditchenko2024whisper}, we extend the training data by combining the LRS3 training set with 1326 hours of English-speaking videos from VoxCeleb2~\cite{Chung18b}. To assess noise robustness, we add babble noise from the MUSAN dataset~\cite{snyder2015musan} at SNRs of 10, 5, 0, $-5$, and $-10$\,dB as in~\cite{rouditchenko2024whisper}. 

We use Whisper Medium~\cite{radford2023robust} as the audio encoder and decoder. Input features are 80-dimensional log-Mel filterbank coefficients extracted from 16\,kHz audio using a 25\,ms analysis window and a 10\,ms frame shift. For the visual modality, frame-level representations are obtained from the AV-HuBERT Large~\cite{shi2022avhubert}. Their weights are frozen during training. As for the LLM, we employ LLaMA2-7B~\cite{touvron2023llama}. A LoRA module is inserted
into each layer of LLaMA with the rank of 16 for fine-tuning. A linear projection layer is introduced to map the visual discrete tokens to the LLaMA embedding dimension.

\begin{table}[t]
		\renewcommand{\arraystretch}{1.2}
		\renewcommand{\tabcolsep}{1.1mm}
		\centering
		\caption{WER (\%) on LRS3 for AVUR-LLM and previous methods in ASR and AVSR under 30\,h, 433\,h, and LRS3+VoxCeleb2 (1759\,h) settings.}
		\resizebox{\columnwidth}{!}{
            \begin{tabular}{ccccc}

				\addlinespace[0.5pt]
				\Xhline{3\arrayrulewidth}
				\textbf{Method}\rule{0pt}{4ex}
				& \makecell{\textbf{Audio}\\\textbf{Encoder}}
				& \makecell{\textbf{Visual}\\\textbf{Encoder}}
				& \makecell{\textbf{Training}\\\textbf{Data (h)}}
				& \textbf{WER(\%)} $\downarrow$\rule[-0.6ex]{0pt}{2.6ex} \\
				\hline
				
				\rowcolor{gray!20}\multicolumn{5}{c}{\textit{Audio-Only}} \\
				CM-seq2seq \cite{ma2021conformer}            & Conformer    & --          & 433        & 2.3 \\
				Fast Conformer \cite{burchi2024multilingual} & Conformer    & --          & 435        & 1.6 \\
				AV-HuBERT \cite{shi2022avhubert}             & Transformer  & --          & 433        & 1.6 \\
				Whisper-finetuned \cite{rouditchenko2024whisper}  & Transformer  & --          & 433        & 2.3 \\
				Auto-AVSR \cite{ma2023auto}                    & Conformer    & --          & 1902/3448  & 1.0/1.0 \\
				Llama-AVSR \cite{Cappellazzo2025Large}                               & Whisper      & --          & 30/433/1759     & 1.5/1.1/0.79 \\
				\hline
				
				\textbf{AVUR-LLM}                                & Whisper      & --          & 30         &  \textbf{1.3} \\
				\textbf{AVUR-LLM}                                & Whisper      & --          & 433        &  \textbf{1.0} \\
                \textbf{AVUR-LLM}                                & Whisper      & --          & 1759        &  \textbf{0.70} \\
				\hline
				\Xhline{1\arrayrulewidth}
				
				\rowcolor{gray!20}\multicolumn{5}{c}{\textit{Audio-Visual}} \\
				CM-seq2seq \cite{ma2021conformer}                       &  \multicolumn{2}{c}{Conformer}        & 433          & 2.3 \\
				AV-HuBERT \cite{shi2022avhubert}                         & \multicolumn{2}{c}{Transformer}       & 433          & 1.4 \\
				CMA \cite{kim2024learning}                         & \multicolumn{2}{c}{Transformer}       & 433          &  1.5 \\
				AV-data2vec \cite{lian2023av}                      & \multicolumn{2}{c}{Transformer}       & 433          & 2.5 \\
				Fast Conformer \cite{burchi2024multilingual}       & \multicolumn{2}{c}{Conformer}         & 1687         & 0.9 \\
				Auto-AVSR \cite{ma2023auto}                        & \multicolumn{2}{c}{Conformer}         & 1902/3448    & 1.0/0.9 \\
				ViT3D-CM \cite{serdyuk2022transformer}             & \multicolumn{2}{c}{Conformer}         & 90K          & 1.6 \\
				LP Conformer \cite{chang2024conformer}             & \multicolumn{2}{c}{Conformer}         & 100K         & 0.9 \\
				Whisper-Flamingo \cite{rouditchenko2024whisper}    & Whisper      & AV-HuBERT              & 433/1759     & 1.1/0.86 \\
				Llama-AVSR \cite{Cappellazzo2025Large}          & Whisper      & AV-HuBERT              & 433/1759          & 0.95/0.77 \\
				Llama-MTSK \cite{cappellazzo2025adaptive}          & Whisper      & AV-HuBERT              & 433          & 0.9 \\
				MMS-LLaMA \cite{yeo2025mms}           & Whisper      & AV-HuBERT              & 433/1759          & 0.9/0.72 \\
				\hline
			
				\textbf{AVUR-LLM}                                & Whisper      & AV-HuBERT              & 433         &  \textbf{0.75} \\
                \textbf{AVUR-LLM}                                & Whisper      & AV-HuBERT              & 1759         &  \textbf{0.68} \\
				\Xhline{3\arrayrulewidth}
		\end{tabular}}
		\label{tab:main}
		\vspace{-0.5cm}
	\end{table}

\subsection{Training and Inference Details}

\noindent\textbf{SMA and AMF Training with N-best Generation.}  
The video data is sampled at 25\,fps and converted to grayscale. Lip-centered regions of size $96 \times 96$ are extracted using Dlib and aligned to a mean face template. During training, random $88 \times 88$ crops are applied with horizontal flipping probability 0.5, while center crops are used during inference. 
For visual discrete tokens, we quantize frame-level features from the 12th AV-HuBERT layer into 2000 K-means clusters, followed by compression~\cite{kim2024efficient}. AdamW optimizer is employed for training with a cosine annealing learning rate scheduler and weight decay of 0.1. The learning rate is set to $1\times10^{-4}$ for training.

\noindent\textbf{VUR with LLM.}  
The N-best hypotheses are paired with the visual discrete tokens to form the input for LLM. The prompt is defined as: \texttt{[Instruction]}: You are an AVSR hypothesis evaluator. Given the input below, assign scores to all candidates based on lip-motion plausibility and linguistic coherence. Input: \texttt{[Visual Units]} \{\textbf{tokens}\} \texttt{[Candidates]} \{\textbf{$y^{(1)}$}, \textbf{$y^{(2)}$}, $\ldots$, \textbf{$y^{(N)}$}\}, where \textbf{tokens} denotes the sequence of visual discrete tokens and \textbf{$y^{(1)}, \ldots, y^{(N)}$} are the $N$ candidate transcriptions from the N-best list. 
LLaMA2-7B is adapted using LoRA fine-tuning with rank 16, a dropout rate of 0.05, and modifications applied to the query, key, value, and output projection layers. Training is performed with a learning rate of $5\times10^{-4}$. Model performance is evaluated using Word Error Rate (WER) for both ASR and AVSR tasks.

\subsection{Experimental Results and Analysis}
\subsubsection{Main Results} 
Table~\ref{tab:main} shows the performance comparisons of the proposed method with prior ASR and AVSR methods. In the audio-only setting, AVUR-LLM attains WERs of 1.3\% (30\,h setting) and 1.0\% (433\,h setting), outperforming Llama-AVSR~\cite{Cappellazzo2025Large} (1.5\%/1.1\%) and Whisper-Flamingo~\cite{rouditchenko2024whisper} (2.3\% at 433\,h setting). Under the extended supervision setting (LRS3+VoxCeleb2, 1759\,h), AVUR-LLM further reduces the audio-only WER to 0.70\%, outperforming Llama-AVSR (0.79\%) at the same scale. In the audio-visual setting, AVUR-LLM attains a WER of 0.75\% with LRS3 (433\,h setting), surpassing recent LLM-based systems~\cite{Cappellazzo2025Large,yeo2025mms,cappellazzo2025adaptive} and establishing a new state of the art among LLM-based AVSR models, outperforming Whisper-Flamingo (1.1\%) and the best prior LLM-based result (0.9\%) by 31.8\% and 16.7\% relative WER reductions, respectively. 
Under the 1759\,h setting, AVUR-LLM achieves 0.68\% WER, consistently outperforming prior LLM-based systems. Notably, at the 433\,h setting, our result remains competitive with Conformer/Transformer systems~\cite{burchi2024multilingual} that are trained on larger supervised corpora. 

\subsubsection{Noise Robustness Evaluation}

\begin{table}[t!]
  \renewcommand{\tabcolsep}{1.3mm}
  \centering
  \caption{WER (\%) on LRS3 (433 h setting) under clean and noisy conditions. We report AVSR results at SNRs \{10, 5, 0, $-5$, $-10$\}\,dB. ``--'' indicates not reported.}

  \vspace{0.2em} 
  \resizebox{\columnwidth}{!}{%
    \begin{tabular}{lccccccc}
      \Xhline{1.5pt}
      \addlinespace[3pt]
      \multirow{2}{*}{\textbf{Method}} & \multirow{2}{*}{\textbf{Modality}} & \multirow{2}{*}{\textbf{Clean}} &
      \multicolumn{5}{c}{\textbf{SNR Level (dB)}, \textbf{WER} (\%) $\downarrow$}\\
     \cmidrule(rl){4-8}
      &  &  &  10  & 5  & 0 & -5  & -10 \\
    \midrule
    \addlinespace[3pt]
    AV-HuBERT \cite{shi2022avhubert}
      & \multirow{6}{*}{AV}\hspace{-0.3em}
      & 1.4
      & 2.0
      & 2.6
      & 5.8
      & 16.6
      & 34.9
      \\
    CMA \cite{kim2024learning}
      &
      & 1.5
      & 1.8
      & 2.4
      & 4.4
      & 11.9
      & 25.8
      \\
    Whisper-Flamingo \cite{rouditchenko2024whisper}
      &
      & 1.1
      & 1.4
      & 2.0
      & 6.3
      & 28.0
      & 42.6
      \\
    AVGER \cite{liu2025listening}
      &
      & 1.1
      & --
      & 1.9
      & 3.7
      & 12.3
      & 29.1
      \\
    Llama-AVSR \cite{Cappellazzo2025Large}
      &
      & 0.95
      & --
      & 2.2
      & 4.2
      & 16.9
      & --
      \\
    MMS-LLaMA \cite{yeo2025mms}
      &
      & 0.9
      & --
      & \textbf{1.3}
      & 2.7
      & 7.4
      & --
      \\
    \addlinespace[2pt]

    \hline
    \addlinespace[3pt]

    \textbf{AVUR-LLM}
      & AV
      & \textbf{0.75}
      & \textbf{1.0}
      &  1.4
      & \textbf{1.7}
      & \textbf{6.3}
      & \textbf{12.9}
      \\
    \addlinespace[3pt]

    \Xhline{1.2pt}
  \end{tabular}%
}
\label{tab:noisy}
\vspace{-0.5cm}
\end{table}

In Table~\ref{tab:noisy}, we evaluate the robustness of AVUR-LLM across acoustic noise levels. AVUR-LLM attains the lowest WER in most of the SNR conditions. In particular, AVUR-LLM surpasses MMS-LLaMA at 0\,dB SNR (1.7\% vs.\ 2.7\%). Across noisy conditions (5/0/$-5$\,dB), AVUR-LLM achieves an overall relative WER reduction of 14.7\%. Under relatively clean conditions, decoder acoustic attention remains highly concentrated, producing low uncertainty and limiting visual injection. However, as the SNR decreases, acoustic attention becomes diffuse. The AMF increases the visual contribution, and the error rate is reduced substantially. At $-5$ and $-10$\,dB, AVUR-LLM reduces WER to 6.3\% and 12.9\%, compared with the best prior results of 7.4\% and 25.8\%, indicating that the proposed method is effective under severe noise.

\subsubsection{Ablation Results}
Table~\ref{tab:ablation} shows the results of our ablation study. When all components are enabled, WER is reduced from 1.30\% to 0.75\% (clean) and from 6.70\% to 1.70\% at 0\,dB SNR, corresponding to relative reductions of 42.3\% and 74.6\%, respectively. Removing VUR markedly degrades robustness (1.70\% to 5.50\%), confirming the effectiveness of the refinement module when coupled with confidence-aware fusion. Using AMF+VUR without SMA already yields a large gain (2.10\%), while the SMA-only setting (w/o AMF and VUR) improves little under noise (6.70\%), indicating that confidence-aware fusion and refinement module are the primary drivers of noise robustness. Adding SMA on top of AMF+VUR further reduces the 0\,dB WER from 2.10\% to 1.70\%, suggesting that sparse alignment provides complementary benefits by aligning multimodal features and mitigating cross-modal mismatch.

\subsubsection{Further Analysis}

\begin{figure}[!t]
  \centering
  \includegraphics[width=1\columnwidth]{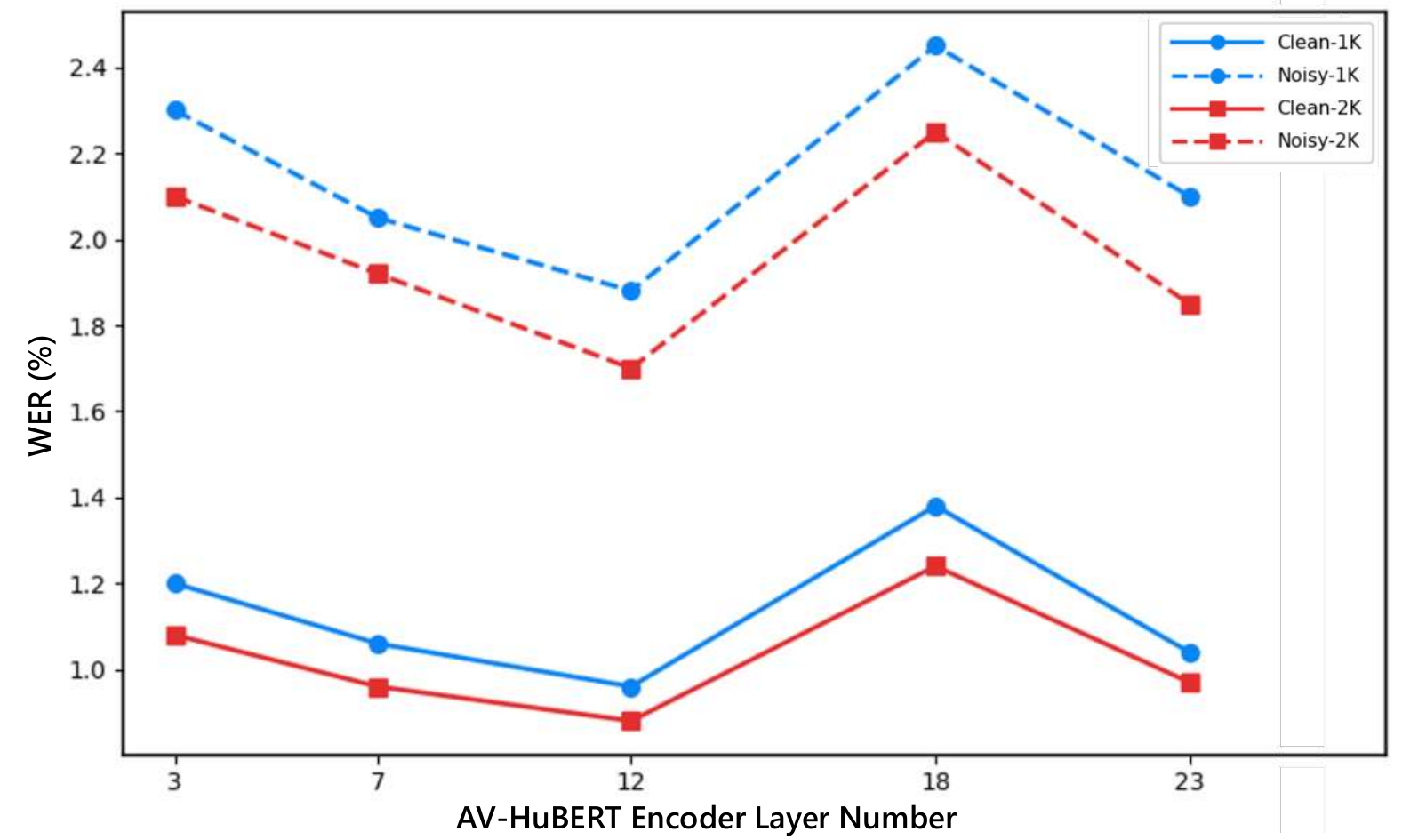}
  \vspace{-6pt}
  \caption{Effect of visual discrete tokens extraction depth and codebook size on WER (\%). Solid lines denote clean, dashed lines denote 0\,dB SNR. Blue curves denote K-means codebook size \(K{=}1000\), red curves denote \(K{=}2000\).}
  \label{fig:fig2}
  \vspace{-8pt}
\end{figure}

Fig.~~\ref{fig:fig2} depicts the effect of visual discrete units extraction depth and codebook size on WER under clean and 0\,dB conditions. Performance peaks at a mid-level layer (12): with 2K clusters, WER reaches $0.75\%$ (clean) and $1.70\%$ ($0$ dB), consistently lower than the 1K setting. Shallow (layer~3) and deep (layer~18) features yield higher error, indicating that mid-level representations balance visual-phonetic detail and invariance. Increasing codebook granularity strengthens VUR’s rescoring by providing more discriminative visual tokens, yielding consistent gains across noise conditions.

\begin{table}[!t]
  \centering
  \caption{Ablation of AVUR-LLM on LRS3 (433 h) under clean and noisy (0\,dB SNR) conditions.}
  \label{tab:ablation}
  \setlength{\tabcolsep}{5pt}
  \renewcommand{\arraystretch}{1.12}
  \resizebox{\columnwidth}{!}{%
    \begin{tabular}{
      l
      c c c
      c c
    }
      \toprule[1.1pt]
      \textbf{Method} & \textbf{SMA} & \textbf{AMF} & \textbf{VUR} & \textbf{Clean} & \textbf{Noisy} \\
      \midrule
      Whisper\nobreakdash-Flamingo~\cite{rouditchenko2024whisper} & -- & -- & -- & 1.10 & 6.30 \\
      Llama-AVSR~\cite{Cappellazzo2025Large}                     & -- & -- & -- & 0.95 & 4.20 \\
      \midrule

      \multicolumn{6}{l}{\textbf{Staged ablation}}\\
      \midrule
      AVUR-LLM (full)              & \cmark & \cmark & \cmark & \textbf{0.75} & \textbf{1.70} \\
      -- w/o VUR                   & \cmark & \cmark & \xmark & 0.97 & 5.50 \\
     -- w/o AMF (SMA only)        & \cmark & \xmark & \xmark & 1.30 & 6.70 \\
      -- w/o SMA (AMF+VUR)         & \xmark & \cmark & \cmark & 0.85 & 2.10 \\
      
      \bottomrule[1.1pt]
    \end{tabular}%
  }
  \vspace{-12pt}
\end{table}

\section{Conclusion}
\label{sec:conclusion}

In this paper, we presented AVUR-LLM, an LLM-based AVSR framework that combines multimodal fusion with visual unit–guided refinement. The method stabilizes cross-modal interaction in the audio pathway, adaptively regulates visual contributions during decoding, and compresses visual features into discrete tokens for rescoring with LLM. A two-stage training scheme preserves the strengths of the pretrained audio and visual encoders while limiting the burden on the LLM.
Experiments demonstrate improvements under both clean and noisy conditions and provide analyses across SNRs and discretization settings, indicating enhanced robustness and data efficiency.

\section{Generative AI Use Disclosure}
Large Language Models (LLMs) were used solely for manuscript polishing (e.g., rephrasing and grammar checks) to improve clarity and readability. The LLMs were not used for ideation, methodology, experimental design, data analysis, or result interpretation. All scientific content was produced and verified by the authors.

\bibliographystyle{IEEEtran}
\bibliography{mybib}

\end{document}